\documentclass[12pt]{iopart}
\usepackage[T1]{fontenc}
\usepackage[latin1]{inputenc}

\begin{document}

\title[Mesoscopic approach to the slow dynamics]{Mesoscopic approach to the slow dynamics of supercooled liquids and colloidal
systems}

\author{A. P\'{e}rez-Madrid, D. Reguera, and J.M. Rub\'{\i}}

\address{Departament de F\'{\i}sica Fonamental, Facultat de F\'{\i}sica, Universitat
de Barcelona,
 Diagonal 647, 08028 Barcelona, Spain}

\begin{abstract}
We propose a method to analyze the dynamics of systems exhibiting slow relaxation
which is based on mesoscopic non-equilibrium thermodynamics. The method allows
us to obtain kinetic equations of the Fokker-Planck type for the probability
functional and their corresponding Langevin equations. Our results are compared
with the ones obtained by other authors. 
\end{abstract}

%\maketitle

\section{Introduction}

The problem of the glass transition has attracted much attention in the past
years, and has been subject of intense experimental, computational and theoretical
investigations. Mode coupling theories and replica techniques have been able
to capture some trends of the phenomenology of these systems. Although interesting
theoretical results have been obtained by means of those theories, the understanding
of the problem is far from being complete \cite{sitges,goetze}.

Supercooled liquids and colloids near a glass transition are characterized by
a extremely slow relaxation of the density variable. As the glass transition
is essentially a dynamic transition, it becomes important to provide a dynamical
description of the evolution of the density variable. In addition, the presence of slow
dynamics in such systems emphasizes the importance of the activated processes,
since they are the main mechanism able to make the system evolve and escape
from its trapping in the metastable states. A complete description of these
systems should then take into account such processes.

In this paper we present a simple mesoscopic formalism able to derive kinetic
equations of the Fokker-Planck type for the probability functional of the density
fields in supercooled liquids and dense colloidal systems. This approach is
based on the hypothesis of local equilibrium in the phase space of the system,
which allows us to describe the processes leading to variations in the state
of these systems by means of non-equilibrium thermodynamics. The extension of
that theory to the mesoscopic level of description has been referred to as mesoscopic
non-equilibrium thermodynamics \cite{prigogine}-\cite{entropic}.

The plan of the paper is as follows. In Section 2, we derive the kinetic equation
for the probability distribution functional of the momentum density and mass
density fields. By adiabatic elimination of the momentum density we obtain the
kinetic equation describing the slow dynamics, giving the configurational changes
in the system. In Section 3, we introduce fluctuations of the distribution functionals
obtaining the corresponding Langevin equation. Finally, in Section 4, we conclude
summarizing our main results.

\section{Kinetic equations}

We consider the system, either a pure liquid or a colloidal suspension, as a
continuum divided in tiny cells of volume \( v_{0} \). In each cell
we define a mass density \( \rho (\bi{r}) \) and a momentum density
\( \bi{g}(\bi{r})=\rho \bi{v} \), with
\( \bi{r} \) being the position vector of a cell and \( \bi{v} \)
the velocity field. For simplicity, we will introduce the compact notation \( \left\{ \underline{\Gamma }\right\} \equiv \left\{ \rho ,\bi{g}\right\}  \).

Let us consider \( \widehat{P}(\left\{ \underline{\Gamma }\right\} ,t) \) as
the distribution functional in phase space, normalized according to the relation
\begin{equation}
\label{normalization}
\int \widehat{P}(\left\{ \underline{\Gamma }\right\} ,t)\delta \underline{\Gamma }=N ,
\end{equation}
 where \( N \) is the number of possible states in phase space. Related to
this distribution functional we introduce the phase space entropy \( S(\widehat{P}(\left\{ \underline{\Gamma }\right\} ,t)) \),
which satisfies the Gibbs equation introduced by mesoscopic non-equilibrium
thermodynamics \cite{degroot}\begin{equation}
\label{gibbs equation}
\bigtriangleup S=-\frac{1}{T}\int \mu (\left\{ \underline{\Gamma }\right\} ,t)\bigtriangleup \widehat{P}(\left\{ \underline{\Gamma }\right\} ,t)\delta \underline{\Gamma }.
\end{equation}
 Here \( T \) is the temperature, and \( \mu  \) represents a chemical potential
defined in phase space satisfying \begin{equation}
\label{chemical potential}
\mu =\frac{\delta S}{\delta \widehat{P}}.
\end{equation}
 Its expression can be obtained with the help of the Gibbs entropy postulate
\cite{degroot}\begin{equation}
\label{gibbs entropy}
S=-k\int \widehat{P}\ln \frac{\widehat{P}}{\widehat{P}^{l.eq.}}\delta \underline{\Gamma }+S^{l.eq.},
\end{equation}
 where \( k \) is the Boltzmann constant. In this expression \begin{equation}
\label{equilibrium distribution}
\widehat{P}^{l.eq.}=\exp \left\{ -\beta \left[ \mu _{o}+H_{K}\left\{ \underline{\Gamma }\right\} +H_{U}\left\{ \rho \right\} \right] \right\} 
\end{equation}
 is the distribution functional at local equilibrium, with \( \beta =1/kT \),
and \( \mu _{o} \) the chemical potential at local equilibrium. Moreover, we
have defined the kinetic energy functional \begin{equation}
\label{kinetic energy}
H_{K}\left\{ \rho ,\bi{g}\right\} =\frac{1}{2}\int \frac{\bi{g}(\bi{r})^{2}}{\rho (\bi{r})}d\bi{r},
\end{equation}
 and the potential energy functional \begin{equation}
\label{potential energy}
H_{U}\left\{ \rho \right\} =\int f\left\{ \rho \right\} d\bi{r}
\end{equation}
 where \( f\left\{ \rho \right\}  \) is the free energy density functional.

By taking variations in eq. ( \ref{gibbs entropy}) we can obtain \begin{equation}
\label{gibbs entropy variations}
\bigtriangleup S=-k\int \bigtriangleup \widehat{P}\ln \frac{\widehat{P}}{\widehat{P}^{l.eq.}}\delta \underline{\Gamma }+\bigtriangleup S^{l.eq.},
\end{equation}
 where \begin{equation}
\label{equilibrium entropy}
\bigtriangleup S^{l.eq.}=-\frac{1}{T}\mu _{o}\bigtriangleup N=-\frac{1}{T}\int \mu _{o}\bigtriangleup \widehat{P}\delta \underline{\Gamma }.
\end{equation}
 Thus, after comparison of eqs. (\ref{gibbs equation}) and (\ref{gibbs entropy variations})
we can infer the expression of the chemical potential \begin{equation}
\label{chemical potential expresion}
\mu =\mu _{o}+kT\ln \frac{\widehat{P}}{\widehat{P}^{l.eq.}}.
\end{equation}

We now assume that the distribution functional evolves according to the continuity
equation \begin{equation}
\label{continuity equation}
\frac{\partial \widehat{P}}{\partial t}+\int \left( \frac{\delta }{\delta \rho }\dot{\rho }\widehat{P}+\frac{\delta }{\delta \bi{g}}\cdot \dot{\bi{g}}\widehat{P}\right) d\bi{r}=-\int \frac{\delta }{\delta \bi{g}}\cdot \bi{J}_{g}d\bi{r},
\end{equation}
 where the dot over the field variables indicates time derivative and \( \bi{J_{g}} \)
is a diffusive current in momentum space.

>From eqs. (\ref{gibbs equation}) and (\ref{continuity equation}) we can obtain
the rate of the entropy variation that is written as

\begin{equation}
\label{rate of entropy}
\frac{\partial S}{\partial t}=\frac{\partial _{e}S}{\partial t}+\sum .
\end{equation}
 In this expression, we can identify the first term on the right hand side as
the rate at which entropy is supplied to the system by its surroundings through
the external constrains, given by \begin{equation}
\label{entropy flux}
\frac{\partial _{e}S}{\partial t}=\frac{1}{T}\int \int \mu \left( \frac{\delta }{\delta \rho }\dot{\rho }\widehat{P}+\frac{\delta }{\delta \bi{g}}\cdot \dot{\bi{g}}\widehat{P}\right) d\bi{r}\delta \underline{\Gamma }.
\end{equation}
 The second contribution corresponds to the entropy produced inside the system
due to the irreversible processes, whose value is \begin{equation}
\label{entropy production}
\sum =-\frac{1}{T}\int \int \bi{J}_{g}\cdot \frac{\delta \mu }{\delta \bi{g}}\delta \underline{\Gamma }d\bi{r},
\end{equation}
where a partial integration has been performed. This entropy production has
the usual form of flux-force pairs from which we can infer the phenomenological
relation \begin{equation}
\label{phenomenological}
\bi{J}_{g}=-\frac{1}{T}\int \underline{L}(\bi{r},\bi{r}^{\prime })\cdot \frac{\delta \mu }{\delta \bi{g}}d\bi{r}^{\prime },
\end{equation}
 where the Onsager coefficients \( \underline{L} \) satisfy the Onsager relations
\( \underline{L}(\bi{r},\bi{r}^{\prime })=\underline{L}(\bi{r}^{\prime },\bi{r})^{\dagger } \),
in which \( ^{\dagger } \) stands for the Hermitian conjugated. By computing
the functional derivative and assuming locality in the coordinates, i.e. \( \underline{L}(\bi{r},\bi{r}^{\prime })=\underline{L}(\bi{r})\delta (\bi{r}-\bi{r}^{\prime }) \),
one obtains \begin{equation}
\label{flow}
\bi{J}_{g}=-\underline{H}(\bi{r})\cdot \left( \beta ^{-1}\frac{\delta }{\delta \bi{g}}+\frac{\delta H_{K}}{\delta \bi{g}}\right) \widehat{P},
\end{equation}
 where \( \underline{H}(\bi{r})\equiv \underline{L}(\bi{r})/T \)
\( \widehat{P} \) can be interpreted as a mobility tensor. Its expression follows
from the Navier-Stokes equation \cite{mazenko}\begin{equation}
\label{mobility}
\underline{H}(\bi{r})=-\eta \left( \frac{1}{3}\bigtriangledown \bigtriangledown +\underline{1}\bigtriangledown ^{2}\right) -\xi \bigtriangledown \bigtriangledown 
\end{equation}
 with \( \eta  \) and \( \xi  \) being the shear and bulk viscosities, respectively,
and \( \underline{1} \) the unit tensor.

By introducing the current we have obtained in eq. (\ref{flow}) into the continuity
equation (\ref{continuity equation}) and using the expressions of \( \dot{\rho } \)
and \( \dot{\bi{g}} \) given by the reversible part of the Navier-Stokes
equation \cite{mazenko}, this yields 

\begin{eqnarray}
\label{fokker-planck}
\fl \frac{\partial \widehat{P}}{\partial t}=\int \left\{ \frac{\delta }{\delta \rho }\bigtriangledown \cdot \bi{g} +\frac{\delta }{\delta \bi{g}}\cdot \left[ \rho \bigtriangledown \frac{\delta H_{U}}{\delta \rho }+\bigtriangledown \cdot \frac{\bi{g}\bi{g}}{\rho }\right] \right. \nonumber\\
+ \left. \frac{\delta }{\delta \bi{g}}\cdot \underline{H}(\bi{r})\cdot \left[ \beta ^{-1}\frac{\delta }{\delta \bi{g}}+\frac{\delta H_{K}}{\delta \bi{g}}\right] \right\} \widehat{P}d\bi{r},
\end{eqnarray}
 which constitutes the functional Fokker-Planck equation for the probability
distribution functional \( \widehat{P}(\left\{ \underline{\Gamma }\right\} ,t) \).

In the diffusive regime, when equilibration in momentum has occurred and only
configurational changes are possible, this equation simplifies considerably.
To achieve the corresponding equation we define the reduced distribution functional
\begin{equation}
\label{reduced probability}
P(\left\{ \rho \right\} ,t)\equiv \int \widehat{P}\delta \bi{g}
\end{equation}
 and the current \begin{equation}
\label{first moment}
\bi{J_{\rho }}\equiv \int \bi{g}\widehat{P}\delta \bi{g}.
\end{equation}
 By applying the time derivative to eq. (\ref{reduced probability}) and using
eq. (\ref{fokker-planck}) we obtain \begin{equation}
\label{reduced continuity}
\frac{\partial P}{\partial t}=\int \frac{\delta }{\delta \rho }\bigtriangledown \cdot \bi{J_{\rho }}d\bi{r}.
\end{equation}
 The probability current defined through eq. (\ref{first moment}) evolves according
to
 \begin{eqnarray}
\label{first moment intermediate}
\fl \frac{\partial \bi{J_{\rho }}}{\partial t}  =  \int \bi{g}\frac{\partial \widehat{P}}{\partial t}\delta \bi{g}\nonumber \\
\fl \mbox{\;\;\;\;\;\;} =  \int \int \frac{\delta \widehat{P}}{\delta \rho (\bi{r}^{\prime })}\bi{g}(\bi{r})\bigtriangledown \cdot \bi{g}(\bi{r}^{\prime })\delta \bi{g}d\bi{r}-\int \widehat{P}\bigtriangledown \cdot \frac{\bi{g}\bi{g}}{\rho }\delta \bi{g}-\rho P\bigtriangledown \frac{\delta H_{U}}{\delta \rho }-\frac{1}{\rho }\underline{H}\cdot \bi{J_{\rho }},
\end{eqnarray}
 where we have used eq. (\ref{fokker-planck}), and partial integration have
been performed taking into account that \( \delta \bi{g}(\bi{r})/\delta \bi{g}(\bi{r}^{\prime })=\delta (\bi{r}-\bi{r}^{\prime }) \).
We now assume that \( \frac{1}{\rho }\underline{H}\cdot \bi{J_{\rho }}\simeq \tau ^{-1}\bi{J_{\rho }} \),
where \( \tau  \) is the characteristic time scale of the inertial regime.
In the diffusive regime when \( \tau \ll t \), assuming equilibration in momentum
we can write \cite{kawasaki}\begin{equation}
\label{adiabatic elimination}
\widehat{P}=\Phi \left\{ \underline{\Gamma }\right\} P(\left\{ \rho \right\} ,t),
\end{equation}
 where \begin{equation}
\label{maxwellian}
\Phi =Z\left\{ \rho \right\} ^{-1}\exp \left( -\beta /2\int \frac{\bigtriangleup \bi{g}^{2}}{\rho }d\bi{r}\right) 
\end{equation}
 is a local Maxwellian with \( \bigtriangleup \bi{g}=\bi{g}-\langle \bi{g}\rangle _{\rho } \),
being \( \langle \bi{g}\rangle _{\rho }=\frac{1}{P}\int \bi{g}\widehat{P}\delta \bi{g} \)
the conditional average, and \begin{equation}
\label{partition function}
Z\left\{ \rho \right\} =Z_{o}\exp \left\{ \frac{3}{2v_{o}}\int \ln \left[ \frac{\rho (\bi{r})}{\rho _{o}}\right] d\bi{r}\right\} ,
\end{equation}
 with \( \rho _{o} \) being the uniform mass density of the system. Thus, by
substituting eq. (\ref{adiabatic elimination}) in eq. (\ref{first moment intermediate}),
performing momentum integration and eliminating inertial terms we achieve \cite{rubi}
\begin{equation}
\label{smoluchowski current}
\bi{J_{\rho }}=-\tau \rho (\bi{r})\bigtriangledown \left( kT\frac{\delta }{\delta \rho }+\frac{\delta H}{\delta \rho }\right) P
\end{equation}
 with \begin{equation}
\label{energy functional}
H\left\{ \rho \right\} =H_{U}-kT\ln Z=H_{U}-\frac{3kT}{2v_{o}}\int \ln \left[ \frac{\rho (\bi{r})}{\rho _{o}}\right] d\bi{r}-kT\ln Z_{o},
\end{equation}
 where we have used the result \( \int \bi{g}(\bi{r})\bi{g}(\bi{r}^{\prime })\Phi \delta \bi{g}=kT\underline{1}\rho (\bi{r})\delta (\bi{r}-\bi{r}^{\prime }) \).
Finally, by substitution of the current \( \bi{J_{\rho }} \) ,
given in eq. (\ref{smoluchowski current}) into eq. (\ref{reduced continuity})
we obtain \begin{equation}
\label{smoluchowski equation}
\frac{\partial P}{\partial t}=-\tau kT\int \frac{\delta }{\delta \rho (\bi{r})}\left[ \bigtriangledown \cdot \rho (\bi{r})\bigtriangledown \right] \left\{ \frac{\delta }{\delta \rho (\bi{r})}+\beta \frac{\delta H}{\delta \rho (\bi{r})}\right\} Pd\bi{r},
\end{equation}
 which constitutes the functional Fokker-Planck equation in the diffusive regime,
in which \( \tau kT \) plays the role of a diffusion coefficient. The stationary
solution is the Boltzmann distribution functional \( P_{st}\sim \exp \left( -\beta H\right)  \)
\cite{frusawa}, thus satisfying detailed balance. This equation has been previously
obtained in \cite{kawasaki} by means of projection operator techniques. It
should be mentioned, as already been pointed out \cite{kawasaki}, that the
presence of a diffusive term in this equation incorporates the existence of
hopping or activated processes.

\section{Fluctuating kinetic equations}

Our purpose in this section is to study the dynamics of the fluctuations in
the distribution functional around a given reference distribution. To this purpose,
we will consider that the distribution solution of eq. (\ref{fokker-planck})
corresponds to the average value over an initial distribution in phase space.
Consequently, the actual value of the distribution functional will differ from
the solution of eq. (\ref{fokker-planck}) in the presence of fluctuations.

In our derivation of the Fokker-Planck equation, we will incorporate such fluctuations by applying fluctuating hydrodynamics in phase space \cite{agusti}. Thus, we
will split up the current \( \bi{J}_{g} \) into systematic and
random contributions \begin{equation}
\label{random current}
\bi{J}_{g}=\bi{J}_{g}^{S}+\bi{J}_{g}^{R}.
\end{equation}
 The former is given precisely by the linear law (\ref{phenomenological}) whereas
the latter defines a Gaussian white noise stochastic process of zero mean and
fluctuation-dissipation theorem given by \begin{equation}
\label{variance}
\fl \langle \bi{J}_{g}^{R}(\left\{ \underline{\Gamma }\right\} ,\bi{r},t)\bi{J}_{g}^{R}(\left\{ \underline{\Gamma }^{\prime}\right\} ,\bi{r}^{\prime },t^{\prime })\rangle _{\widehat{P}_{o}}=2k\underline{L}\delta (\left\{ \underline{\Gamma }\right\} -\left\{ \underline{\Gamma }^{\prime}\right\} )\delta (\bi{r}-\bi{r}^{\prime })\delta (t-t^{^{\prime }}),
\end{equation}
 where \( \underline{L} \) has been defined in eq. (\ref{phenomenological}),
and \( \widehat{P}_{o} \) is an initial probability distribution.

When employing the decomposition (\ref{random current}) in the continuity equation
(\ref{continuity equation}), we obtain \begin{equation}
\label{fluctuating continuity}
\frac{\partial \widehat{P}^{*}}{\partial t}=-\int \left( \frac{\delta }{\delta \rho }\dot{\rho }\widehat{P}^{*}+\frac{\delta }{\delta \bi{g}}\cdot \dot{\bi{g}}\widehat{P}^{*}+\frac{\delta }{\delta \bi{g}}\cdot \bi{J}_{g}^{S}+\frac{\delta }{\delta \bi{g}}\cdot \bi{J}_{g}^{R}\right) d\bi{r},
\end{equation}
 where \( \widehat{P}^{*} \) is the fluctuating probability distribution. This
equation can also be expressed as 

\begin{eqnarray}
\label{fluctuating fokker-planck}
\fl \frac{\partial \widehat{P}^{*}}{\partial t}=\int \left\{ \frac{\delta }{\delta \rho }\bigtriangledown \cdot \bi{g}+\frac{\delta }{\delta \bi{g}}\cdot \left[ \rho \bigtriangledown \frac{\delta H_{U}}{\delta \rho }+\bigtriangledown \cdot \frac{\bi{g}\bi{g}}{\rho }\right] \right. \nonumber \\
\lo+ \left. \frac{\delta }{\delta \bi{g}}\cdot \underline{H}(\bi{r})\cdot \left[ \beta ^{-1}\frac{\delta }{\delta \bi{g}}+\frac{\delta H_{K}}{\delta \bi{g}}\right] \right\} \widehat{P}^{*}d\bi{r}+\int \frac{\delta }{\delta \bi{g}}\cdot \bi{J}_{g}^{R}d\bi{r},
\end{eqnarray}
 which corresponds to the fluctuating functional Fokker-Planck equation.

Using the same adiabatic elimination procedure performed in the previous section
we can obtain the fluctuating functional Fokker-Planck equation in the diffusive
regime, yielding 

\begin{equation}
\label{fluctuating smoluchowski}
\fl \frac{\partial P^{*}}{\partial t}=-\tau kT\int \frac{\delta }{\delta \rho (\bi{r})}\left[ \bigtriangledown \cdot \rho (\bi{r})\bigtriangledown \right] \left\{ \frac{\delta }{\delta \rho (\bi{r})}+\beta \frac{\delta H}{\delta \rho (\bi{r})}\right\} P^{*}d\bi{r}+\tau \int \frac{\delta }{\delta \rho }\bigtriangledown \cdot \int \bi{J}_{g}^{R}\delta \bi{g}d\bi{r}.
\end{equation}

>From this equation it is possible to get the Langevin equation for the first
moment of the distribution functional, defined through \begin{equation}
\label{density}
\overline{\rho }(\bi{r},t)=\int \rho (\bi{r})P^{*}(\left\{ \rho \right\} ,t)\delta \rho (\bi{r}).
\end{equation}
 By multiplying eq. (\ref{fluctuating smoluchowski}) by \( \rho (\bi{r}) \)
and integrating one obtains \begin{equation}
\label{langevin}
\frac{\partial \overline{\rho }(\bi{r},t)}{\partial t}=\tau kT\int \left[ \bigtriangledown \cdot \rho (\bi{r})\bigtriangledown \right] \left\{ \frac{\delta }{\delta \rho (\bi{r})}+\beta \frac{\delta H}{\delta \rho (\bi{r})}\right\} P^{*}\delta \rho (\bi{r})+\eta (\bi{r},t),
\end{equation}
 where \( \eta (\bi{r},t)=-\tau \bigtriangledown \cdot \int \bi{J}_{g}^{R}\delta \rho \delta \bi{g} \)
is the noise term, whose correlation, after eq. (\ref{variance}) is used, is
given by \begin{equation}
\label{correlation}
\langle \eta (\bi{r},t)\eta (\bi{r}^{\prime },t^{^{\prime }})\rangle =2kT\tau \bigtriangledown _{r}\cdot \overline{\rho }(\bi{r},t)\bigtriangledown _{r^{\prime }}\delta (\bi{r}-\bi{r}^{\prime })\delta (t-t^{^{\prime }}).
\end{equation}
 The first term inside the bracket in eq. (\ref{langevin}) is the noise-induced
drift \cite{kampen}, which arises from the nonlinear character of the dynamics
which is due to the density dependence of the kinetic coefficient.

Assuming \( P^{*}=\delta \left( \left\{ \rho \right\} -\left\{ \overline{\rho }\right\} \right)  \),
eq. (\ref{langevin}) yields \begin{equation}
\label{langevin2}
\frac{\partial \overline{\rho }\left( \bi{r},t\right) }{\partial t}=\tau \bigtriangledown \cdot \overline{\rho }(\bi{r},t)\bigtriangledown \frac{\delta H\left( \left\{ \overline{\rho }(\bi{r},t)\right\} \right) }{\delta \overline{\rho }(\bi{r},t)}+\eta (\bi{r},t)\, 
\end{equation}
 which coincides with the stochastic equation in terms of the number density
field \( \rho \left( \bi{r},t\right)  \) in Refs. \cite{frusawa},
\cite{dean}-\cite{tanaka}. The average over the noise of eq. (\ref{langevin2})
yields the mean field equation proposed in Ref. \cite{tarazona}

\section{Conclusions}

In this paper, we have proposed a method to obtain Fokker-Planck equations for
the probability distribution functional of the density fields for a liquid in
the framework of mesoscopic non-equilibrium thermodynamics \cite{rubi}. Keeping
the essentials of non-equilibrium thermodynamics, and extending its range of
validity to the mesoscopic domain we have proposed a Gibbs equation where the
entropy depends on the probability distribution functional in phase space. The
peculiar dynamics of glassy systems enables one to assume the existence of two
dynamical regimes, one related to the fast processes or inertial regime, the
other related to the slow relaxation processes or diffusive regime. After, adiabatic
elimination of the fast degrees of freedom \cite{rubi}, we have obtained the
Fokker-Planck equation for the probability distribution functional of the fluid
mass density, which describes the slow dynamics of the system. The previous
description is independent of the underlying microscopic model. The important
requirement is the existence of well separated time and length scales in the
system, which sustain the validity of the local equilibrium hypothesis in phase
space. Supercooled liquids and dense colloidal suspensions are known to satisfy
this requirement.

At mesoscopic level we have considered the hydrodynamic fields of the liquid
as stochastic variables whose deterministic dynamic is governed by the Navier-Stokes
equations. From these equations we can identify the phenomenological coefficients
in the kinetic equations. We have also analyzed fluctuations of the distribution
functional by adding a stochastic contribution to the current in phase space
which satisfies a fluctuation-dissipation theorem. From this procedure we obtain
a Langevin equation for the mass density.

One of the advantages of the formalism we have proposed in this paper is that
the dynamic description in terms of a probability functional provides a complete
information of the slow dynamics of the system. In addition, its validity is
not restricted to the particular situations addressed in this paper. The same
kind of description is perfectly applicable to a wide variety of different situations,
among which we can mention the cases of granular materials \cite{morgado} or
nucleation and phase transitions in inhomogeneous media \cite{oxtoby,bray}.

\ack This work has been partially supported by DGICYT of the
Spanish Government under grant PB98-1258. D. Reguera wishes to thank Generalitat de Catalunya
for financial support.

\end{document}